\documentclass[aps,11pt,prd,superscriptaddress,preprintnumbers, 
	amsfont,
	amssymb,
	amsmath,
	notitlepage,
	longbibliography,
	nofootinbib]{revtex4-1}
\usepackage[colorlinks=true, 
	linkcolor=blue, 
	citecolor=blue, 
	urlcolor=blue, 
	linktocpage=true]{hyperref}
\usepackage[english]{babel}
\usepackage{graphicx}

\begin{document}

\title{Dark matter droplets}
\date{\today} 

\author{Ian G. Moss}
\email{ian.moss@newcastle.ac.uk}
\affiliation{School of Mathematics, Statistics and Physics, Newcastle University, 
Newcastle Upon Tyne, NE1 7RU, UK}

\begin{abstract}
A new model for dark matter is put forward which consists of uniform droplets of
Bose Einstein condensate. In this model, structure forms rapidly,  shortly after
the hot big bang plasma de-ionises. The model also produces modifications to the expansion 
rate before droplet formation that affect the measurement of cosmological parameters
from Cosmic Microwave Background data. The model could contribute to explaining
why observations at high redshift see anomalously high structure formation and
predict low values for the  Hubble constant.
\end{abstract}

\maketitle

\noindent{\it Introduction.} Amongst the many ideas about the nature of dark matter is the suggestion
that it consists of oscillations in a very light scalar field 
\cite{Preskill:1982cy,Abbott:1982af,Dine:1982ah,PhysRevLett.85.1158,Schive:2014dra,PhysRevLett.85.1158}
 (see \cite{Marsh:2015xka,Hui:2021tkt,Ferreira:2020fam} for reviews).
Generally, scalar field dark matter models go under the name of ``fuzzy dark matter", because the matter
clusters in a similar way to particulate dark matter,  but with some wavelike interference
superimposed.  In scalar field models that include self-interaction,  structures involve a balance of scalar field forces,
which can be attractive or repulsive,  spatial gradient pressure and gravitational forces
\cite{Chavanis:2011zi,Chavanis:2011zm,Desjacques:2017fmf,Chavanis:2020rdo,PhysRevD.109.103518}.
These models also lead to modifications to the equation of state,  beyond partuculate dark matter,
and thereby affect the expansion history of the universe \cite{Johnson:2008se}. 
In this paper, a new type of fuzzy dark matter model is introduced,  in which the self-interaction switches from being 
repulsive at early times, to attractive at later times. 

Self-interacting fuzzy dark matter models can form structures even in the absence of
gravity.  For real scalar fields, these structures are localised,  but oscillatory in time,  and therefore 
known as ``oscillons'' \cite{Makhankov:1979fc,Gleiser:1991rf,Gleiser:1993pt}.
They can form from an instability in the field theory \cite{Johnson:2008se},  and have mostly been studied 
in the context of reheating of the early post-inflationary universe \cite{Amin:2010xe,Amin:2010dc}.
Oscillons are long-lived,  but ultimately decay.  Charged fields,  on the other hand,  have conserved charges
and can give rise to stable configurations that can persist over cosmological timescales.

The fuzzy dark matter model introduced here can be placed in the context of a Coleman-Weinberg potential
inspired by elementary particle physics \cite{PhysRevD.7.1888}. 
The switch in self-interaction closely parallels some laboratory experiments on cold atom mixtures,
where droplets form as a result of an instability that can be triggered by quantum vacuum polarisation effects
\cite{PhysRevLett.115.155302,PhysRevLett.122.090401,Cabrera_2018,Semeghini_2018,Cheiney_2018,D_Errico_2019,PhysRevResearch.3.033247}. 
The droplets can also be viewed as a type of oscillon solution with a flat density profile \cite{Amin:2010jq}.
Droplets form rather like liquid water droplets form in clouds. Merging droplets can form 
larger drops, which can start to feel the effects of gravity, the Earth's gravity in the case of clouds 
or self-gravity in the case of cosmology.

The quantum droplets have uniform density, surrounded by vacuum. This sets them apart from other types of
droplet which balance quantum pressure and gravity 
\cite{Chavanis:2011zi,Chavanis:2011zm,Desjacques:2017fmf,Chavanis:2020rdo,Delgado:2022vnt,PhysRevD.109.103518}.
They can range in mass from microscopic, to many solar masses.  A major feature of the quantum droplets
is that they can grow exponentially in time \cite{Johnson:2008se},  unlike the linear growth typical of
gravitational instability.  Baryons found inside the droplets would respond to the enhanced density.
This means they can form structures very early on in the history of the universe from very
small initial perturbations. Quantum droplets are ideally situated to play a role in explaining how 
star and black hole formation seems to have started at such large redshifts \cite{Finkelstein_2022,Labb__2023}.

Modifications to the dark matter sector at high redshift can also affect predictions of the cosmological parameters
based on Cosmic Microwave Background (CMB) data. This can be relevant because the value of Hubble's constant $h$ predicted in 
Lambda Cold Dark Matter ($\Lambda {\rm CDM}$) models \cite{Planck:2018vyg} is currently lower than the 
Hubble constant  measured at low redsfhift \cite{Riess:2016jrr,Riess:2021jrx}. There is
a long list of models that have been put forward to explain this Hubble tension
\cite{DiValentino:2021izs,Schoneberg:2021qvd,Kamionkowski:2022pkx} including some with fuzzy dark matter that 
have used a ``frozen field" \cite{Smith:2019ihp} or dark matter-dark energy interactions \cite{Uzan:2023dsk}. 
In the new model,  Droplet formation at redshift around 10--100 decreases the size of the sound horizon at the surface of last scattering
compared to $\Lambda {\rm CDM}$ models. This increases the CMB 
prediction of the Hubble constant. After the droplets form, they behave as ordinary  $\Lambda {\rm CDM}$.
On the downside, models which decrease the size of the sound
horizon often conflict with other observable parameters \cite{Jedamzik:2020zmd}. For this preliminary
investigation, the effect on the CMB is primarily a constraint on the model, with moves in the right direction
for solving the Hubble tension problem.

The cosmology of Coleman-Weinberg based potentials have been considered previously
as a means of generating particle masses \cite{Steele:2013fka},  and in the context
of relic abundances in extensions of the standard model of particle physics \cite{Murgui:2023kig}.
The model considered here has exceptionally small masses and couplings. This aspect of the model is not 
fully justified,  but is common to most fuzzy dark matter models.

 \noindent{\it Coleman-Weinberg models}--
The dark matter models we will investigate have a transition from repulsive to attractive self interactions.
This can be realised in scalar fields with a Coleman-Weinberg effective potential.
Consider a charged scalar
field $\chi$ that interacts with an (invisible) photon. We require that the scalar self-interaction
constant $\lambda$ and scalar mass $m$ are both small, and then the Lagrangian density ${\cal L}$ is
\begin{equation}
{\cal L}=\frac{1}{2c^2}|\dot\chi|^2-\frac12|\nabla \chi|^2-\frac12\frac{m^2c^2}{\hbar^2}|\chi|^2
-\frac34\frac{\alpha^2}{\hbar c}|\chi|^4\ln\frac{|\chi|^2}{\mu^2_R},
\end{equation}
where $\mu_R$ is a renormalisation scale. The logarithmic term represents the effect of vacuum polarization.
The parameter requirements for the vacuum polarization to be important are $\lambda\ll \alpha^2$, 
where $\alpha$ is fine structure constant for the
scalar-photon interaction, and $\lambda m\ll \hbar \alpha^2|\chi|^2/c^3$.

The field equations in an expanding flat universe with scale factor $a(t)$ can be derived from the Lagrangian
given above. In the non-relativistic limit, used widely for scalar dark matter, we introduce a complex field
$\psi$ and set
\begin{equation}
\chi=\frac{\hbar}{\sqrt{m a^3}} e^{-im c^2 t/\hbar}e^{-i\theta(t)}\psi\label{psi}
\end{equation}
The reason for the additional
phase $\theta$ should become clearer below. Dropping the $\ddot\psi$ terms, we find
\begin{equation}
i\hbar\dot\psi=-\hbar\dot\theta\psi-\frac{\hbar^2}{2m}\nabla^2\psi+gn\psi\left(\ln\frac{n}{n_d}-\frac12\right)\label{gpe}
\end{equation}
where $n=|\psi|^2/a^3$ is the particle number density, $n_d$ is related to $\mu_R$ and the coupling $g$ is
\begin{equation}
g=\frac32\frac{\hbar^3\alpha^2}{m^2 c}
\end{equation}
Introducing the $\theta$ term allows this equation to have a stationary solution. If we choose
\begin{equation}
\hbar\dot\theta=gn\left(\ln\frac{n}{n_d}-\frac12\right)\label{mu}
\end{equation}
then $\psi$ is constant and the number density decreases as ordinary matter.
In the theory of cold atoms, we would identify $\hbar\dot\theta$ with a chemical potential $\mu$. In
the expanding universe, $\mu$ decreases with time. There are additional corrections to the density redshift relation as
a consequence of the neglected derivative terms, which will be considered later.

\noindent{\it Droplet formation}.--Note that the effective coupling in Eq. (\ref{gpe}) becomes negative when $n<n_d\sqrt{e}$, and we
should expect some type of instability.  This can be analysed using cosmological perturbation theory,
as in Ref \cite{Desjacques:2017fmf}. The density inhomogeneity $\delta=\delta n/n$ for a comoving mode $k$ satisfies
\begin{equation}
\ddot \delta +2H\dot\delta +\left(\frac{\hbar^2 k^4}{4 m^2 a^4}+\frac{gn k^2}{m a^2}\ln\frac{n}{n_d}-4\pi G nm\right)\delta
=0.
\end{equation}
The negative sign in the last term indicates that large wavelengths have the usual Jeans instability, which grows linearly with time. 
Additionally, modes can also become unstable due to the scalar self force when $n<n_d$. The parameter range 
$g n_d/2\hbar\gg H$ is particularly interesting because this is where perturbations can grow faster than the Hubble flow.
There are approximate solutions  $\delta\propto e^{-i\omega t}$,
with dispersion relation depending on the physical wavenumber $k_{\rm phys}=k/a$,
\begin{equation}
\omega^2=\frac{k_{\rm phys}^2}{2m}\left[\frac{\hbar^2k_{\rm phys}^2}{2m}+2gn\ln\frac{n}{n_d}\right]
\end{equation}
When $n<n_d$, imaginary values of $\omega$ are associated with modes that grow exponentially.
The fastest growing mode for given $n$ is the one with the largest value of $|\omega|$,  with growth rate
$v_d$,  at physical wavenumber $k_{\rm max}$,
\begin{equation}
v_d=\frac{\hbar k_{\rm max}^2}{2 m}=\frac{gn}{\hbar}\left|\ln\frac{n}{n_d}\right|.
\label{max}
\end{equation}
The maximum growth rate $v_d$ occurs when $n=n_d/e$, for a length scale
$\lambda_d=2\pi/k_{\rm max}$,
\begin{equation}
\lambda_d=\left(\frac{m c^2}{g n_d}\right)^{1/2}\lambda_{CW},\label{size}
\end{equation}
where $\lambda_{CW}$ is the Compton wavelength of the particle $2\pi\hbar/m c$.
In the chosen parameter range, $|\omega|\gg H$ and the droplet formation is driven by scalar field
and not by gravitational forces.  
Another way of seing this is that in the range $g n_d/2\hbar<H$,  $\lambda_d$ is smaller than the Jeans 
length $\lambda_J$ .
(This can be seen from the relation $2\hbar H/gn_d\sim \lambda_d^2/\lambda_J^2$.)

\begin{center}
\begin{figure}[ht]
\begin{center}
\scalebox{0.3}{\includegraphics{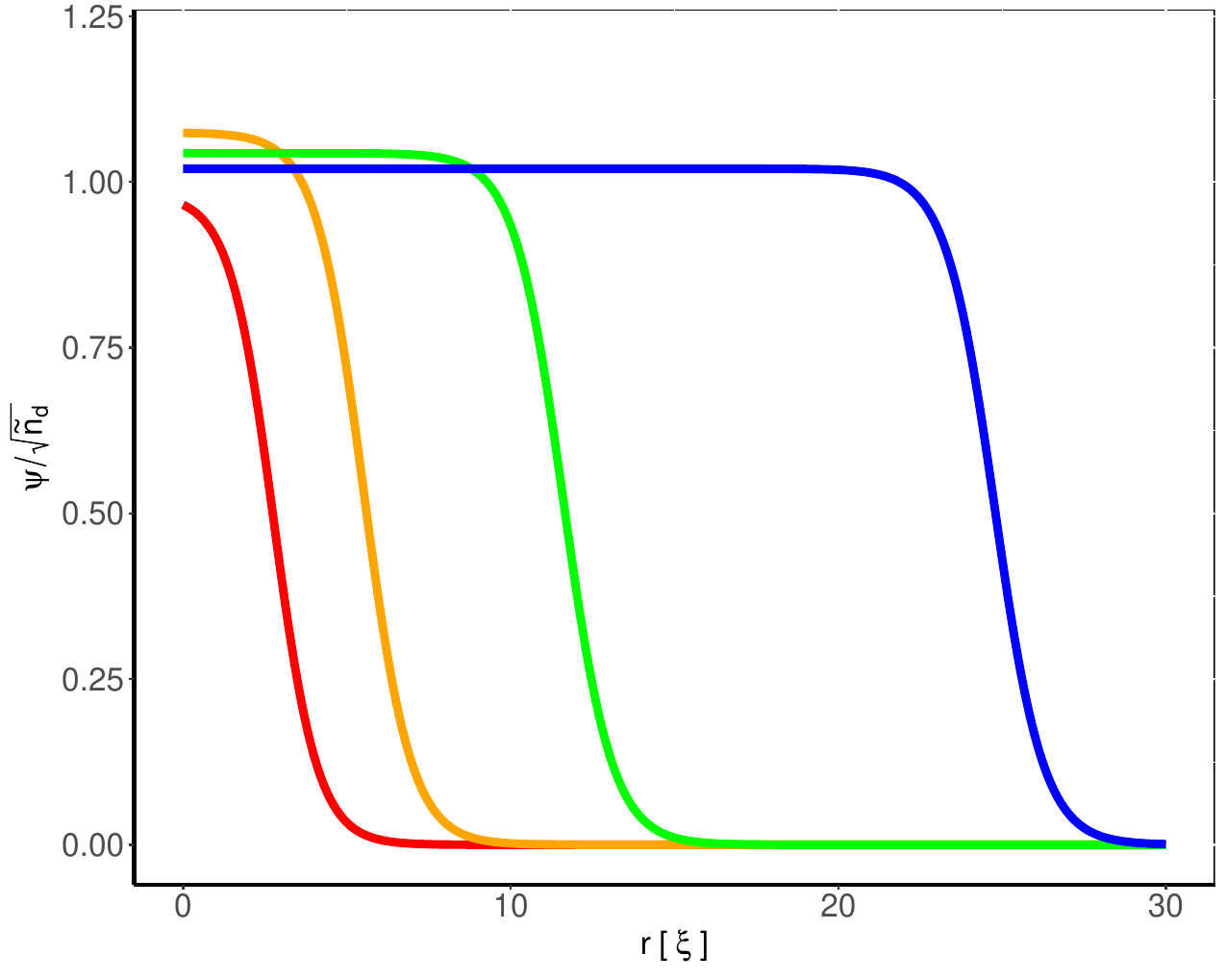}}
\end{center}
\caption{Droplet profiles for a range of particle number from $100n_d\xi^3$ to $10^4n_d\xi^3$,
where the length scale $\xi=\hbar/(mgn_d)^{1/2}$.
The larger droplets have interiors with constant density $n_d\sqrt{e}$ and and edge with thickness of order $\xi.$
}
\label{fig:shape}
\end{figure}
\end{center}

On scales smaller than the Jeans length the system is governed by the GPE (\ref{gpe}) 
which has stable droplet solutions.  The lowest energy state
of an idividual droplet depends on the radius,  or equivalently on the particle number in
the droplet.  Solving the GPE for fixed particle number gives the profiles
shown in figure \ref{fig:shape}.

The numerical solutions above neglected gravity.  Let us include this for spherical droplet of radius $R$, mass
$M$, uniform density $n_R$. and surface tension $\sigma$.  The total energy of the droplet $E_d$ is
\begin{equation}
E_d=\frac12 \frac{gM}{m}n_R\left(\ln\frac{n_R}{n_d}-1\right)-\frac{GM^2}{2R}+\sigma R^2+Mc^2
\end{equation}
For fixed mass, $R\equiv R(n_R)$ and the generalised force $dE_d/dn_R$ is
\begin{equation}
\frac{dE_d}{dn_R}=\frac{Mc^2}{2 n_R}\left\{
\frac{gn_R}{m c^2}\ln\frac{n_R}{n_d}-\frac{GM}{3 c^2R}-\frac{4\sigma R^2}{3 Mc^2}\right\}
\end{equation}
Consideration of the gradient terms gives $\sigma\approx\kappa g n_R^2\lambda_d$, where $\kappa$
is a small numerical constant. The first and the last terms balance for any $R\ge \lambda_d$, 
with $n_R\approx n_d$. When $R<\lambda_J$, the gravity factor is negligible. Uniform droplets
exist for radii $\lambda_d\le R<\lambda_J$.

Once formed, droplets at rest would behave like very large CDM particles.  In practice, droplets can form with a range 
of sizes and initial velocities, and may combine like water droplets into larger drops. Larger drops become non-uniform
as they are are effected increasingly by gravitational and pressure forces. This process could play a role
in early structure formation, but it is quite complex, and we leave for it future work.

\noindent{\it Effect on the CMB}.--Next, we turn to the deviations from the $\epsilon\propto a^{-3}$ 
rule that happen before the droplet phase. The simplest 
approach is to use the conservation rule
\begin{equation}
\dot\epsilon+3H(\epsilon+p)=0\label{cons}
\end{equation}
The energy density 
$\epsilon$ of a homogeneous region is
\begin{equation}
\epsilon=nm c^2+\mu n+\frac12 gn^2\left(\ln\frac{n}{n_d}-1\right)\label{eps}
\end{equation}
A similar calculation gives the pressure
\begin{equation}
p=\mu n-\frac12 gn^2\left(\ln\frac{n}{n_d}-1\right)\label{p}
\end{equation}
Now, we make use of the non-relativistic approximation $m c^2\gg gn$. We replace $n\to n+\delta n$,
in Eqs. (\ref{eps}) and (\ref{p}), where $n$ has the CDM behaviour $n\propto a^{-3}$. Solving for
the number density perturbation $\delta n$ from CDM in Eq. (\ref{cons}) gives
\begin{equation}
\frac{\delta n}{n}=-\frac{gn}{m c^2}\ln\frac{n}{n_d}\label{dn}.
\end{equation}
Substituting back into the energy,
\begin{equation}
\frac{\delta\epsilon}{\epsilon}=\frac12\frac{gn}{m c^2}\left(\ln\frac{n}{n_d}-1\right)\label{deps}
\end{equation}
This only affects the Hubble parameter-redshift relation for redshifts larger than the redshift of droplet formation $z_d$.
If we denote the redshift of matter radiation equality by $z_{\rm eq}$, then the dark matter dominates the expansion
for redshifts $1\ll z_d<z<z_{\rm eq}$, and in this range $H\propto \epsilon^{1/2}$. If we hold the Hubble constant $h$ fixed, then the 
Hubble parameter $H(z_*)$ at the surface of last scattering $z_*$ for the dark matter model and $H_{CDM}(z_*)$
for ordinary CDM are related by
\begin{equation}
\delta \ln H_*\equiv\frac{H(z_*)-H_{CDM}(z_*)}{H_{CDM}(z_*)}\approx \frac12 \frac{\delta \epsilon_*}{\epsilon_*}
\label{dH}
\end{equation}
Using Eq. (\ref{deps}), we have an increase in energy density between redshift $z_d$ and $z_*$,
\begin{equation}
\delta \ln \epsilon_*=\frac12\frac{g n_d}{m c^2}\left(\frac{1+z_*}{1+z_d}\right)^3\left\{\ln\left(\frac{1+z_*}{1+z_d}\right)^3-1 \right\}
\end{equation}
In other words, the Hubble parameter at decoupling would appear larger in the dark matter model than it would when assuming
CDM.

The effect of the dark matter model on CMB observations can be estimated as follows. 
The primary observable quantity is the angle subtended by the sound horizon
at last scattering $r_s(z_*)$, as measured in the position of the first peak in the angular spectrum.
The angle is given by $\theta_s=r_s(z_*)/r_H$, where $r_H$ is the distance to the cosmological
horizon. For $z_d\gg 1$, the main dependence on $\delta\epsilon$ is through
$r_s(z_*)\sim 2c_s/H(z_*)$, because the luminosity distance $r_H$ is determined
mostly at low $z$ where the droplet model is identical to $\Lambda$CDM. We therefore expect from Eq. (\ref{dH}) that
$\delta\ln \theta_s\sim -0.5\,\delta\ln\epsilon_*$.

The predicted value of the Hubble parameter $h$ is 
inferred from a complicated numerical pipeline, but Percival et al. \cite{2dFGRSTeam:2002tzq} 
give some useful approximations relating deviations in the horizon angle $\theta_H$ to the matter density 
parameter $\omega_m=\Omega_m h^2$ and Hubble constant $h$. Applying a similar method, including the Hubble 
law deviation, gives
\begin{equation}
\delta\ln \theta_H\approx 0.2 \,\delta \ln h+0.14\,\delta\ln\omega_m-0.55\,\delta\ln\epsilon_*.
\end{equation}
A secondary effect on the CMB comes through changes to Silk damping of the CMB peaks, 
but the change turns out to be very minor, and the peak heights effectively fix the value of $\omega_m$.
If we fix $\theta_H$ by the observations, then predicted value of the Hubble constant $h$
is larger in a model with droplet formation after the surface of last scattering, by
$\delta\ln h\approx 2.5\,\delta\ln\epsilon_*$.

\begin{center}
\begin{figure}[ht]
\begin{center}
\scalebox{0.35}{\includegraphics{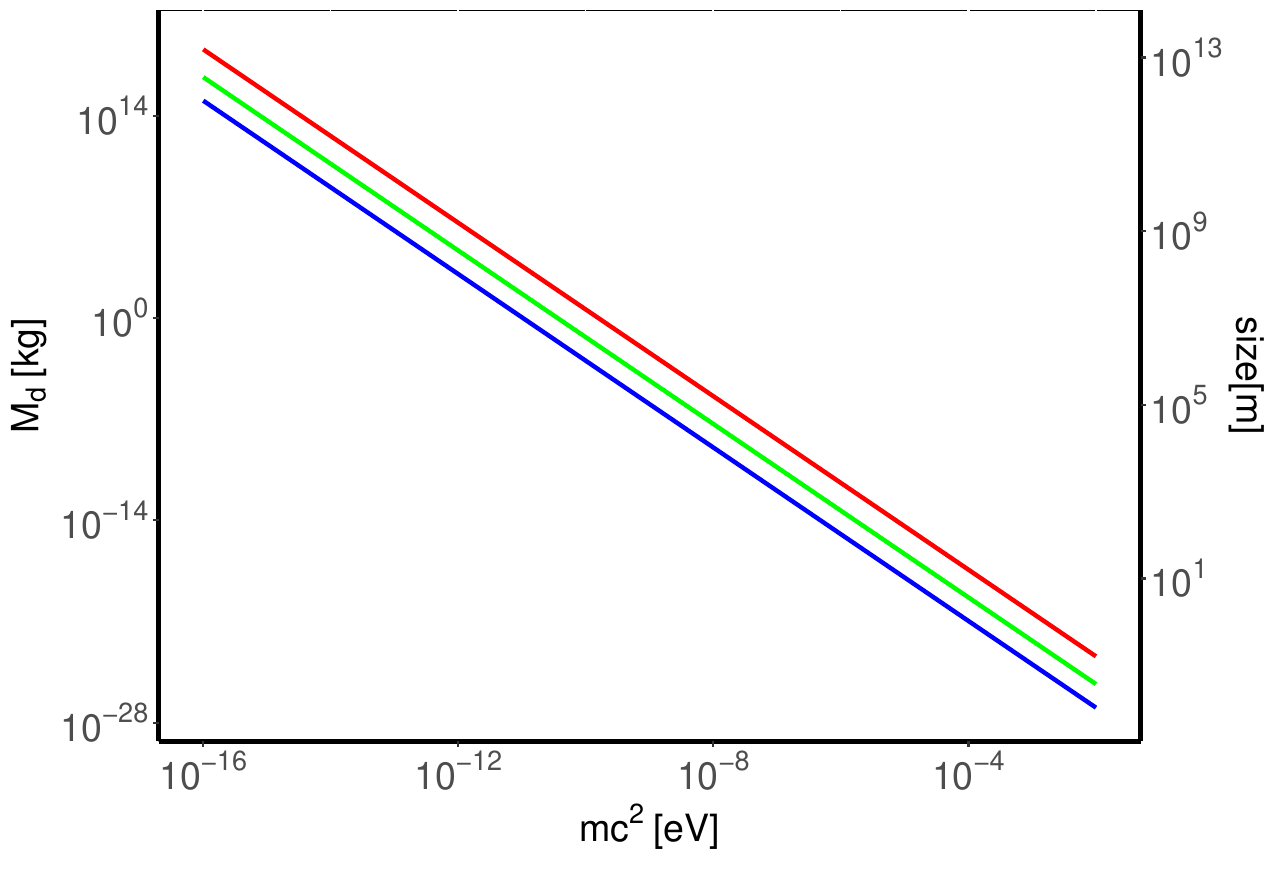}}
\end{center}
\caption{Droplet masses and sizes for a range of dark matter particle masses. The three curves are for
droplets forming at redshift 10 (red),  100 (green) and 400 (blue). The parameters have been chosen so that
$\delta \ln h=5\%$.
}
\label{fig:paras}
\end{figure}
\end{center}

The dark matter model has three parameters, the particle mass $m$, the coupling $g$ (or $\alpha$) 
and the droplet density $n_d$.  The CMB observations are fixed by the redshift of droplet formation 
$z_d$,  and the combination  $gn_d/mc^2$.  This leaves a third free parameter that fixes the 
droplet size.  One constraint on the parameters which we have not discussed so far is that the density 
variation $\delta\ln\epsilon$ must be small at matter-radiation equality. This because the energy 
density behaves like radiation when the $\chi^4$ term starts to dominate,  and nucleosynthesis
constraints kick in.  This limits the change in $h$ to a few percent.
Some illustrative parameters are given in figure \ref{fig:paras}.

\noindent{\it Conclusions} Although the results have been built around a fuzzy dark matter model with Coleman-Weinberg
potential, some of the features are more general. Structure can grow at an exponential rate in dark matter with an 
attractive self-interaction, compared to the usual linear growth from gravitational forces. On the other hand, fuzzy
dark matter with repulsive self-interactions increase the value of the Hubble constant $h$ predicted by the CMB.
The Coleman-Weinberg models combine both, with a self-interaction that is repulsive at large redshift and attractive at
low redshift.

The Coleman-Weinberg dark matter potentials lead to droplet formation, with a wide range of possible droplet masses
and sizes. The simplest fate of these droplets is to remain as dark matter particles.
A more interesting possibility is that the droplets move and merge to form larger drops. They would also be
affected by the gravitational forces from baryonic matter. Understanding the details would require a more
detailed analysis. Through an interesting coincidence, cold atom models in two spatial dimensions 
have the same logarithmic
form for the vacuum polarization as Coleman-Weinberg models in three spatial dimensions 
\cite{PhysRevLett.117.100401}. This opens up
the possibility of making a laboratory analogue of cosmological structure formation, to study the initial size and velocity 
distribution of  droplets when they form, and the subsequent process of merger into larger drops.

\acknowledgments
The author is grateful for discussion with Matt Johnson, Tom Billam and Chanda Prescod-Weinstein.
This work is supported by the UK Science and Technology Facilities Council
[grants ST/T00584X/1 and ST/W006162/1]. The author is also grateful for the hospitality of Perimeter 
Institute where part of this work was carried out.  Research at Perimeter Institute is supported in part 
by the Government of Canada through the Department of Innovation, Science and Economic 
Development and by the Province of Ontario through the Ministry of Colleges and Universities.

\bibliography{paper.bib}

\end{document}